\documentclass{iau}
\usepackage{graphicx}
\usepackage{float}
\usepackage{wrapfig}

\title[Discovery of a primordial water reservoir in the envelope of HH\,211] 
{Discovery of a primordial water reservoir in the envelope of HH\,211}

\author[Odysseas Dionatos]   
{Odysseas Dionatos$^1$
 }

\affiliation{$^1$Department of Astrophysics, University of Vienna, T{\"u}rkenschanzstrasse. 17, 1180 Vienna, Austria \\ email: {\tt odysseas.dionatos@univie.ac.at} \\[\affilskip]
}

\pubyear{2019}
\volume{345}  
\jname{Origins: from the Protosun to the First Steps of Life}
\editors{Bruce G. Elmegreen, L. Viktor T\'oth, Manuel G\"udel, eds.}
\begin{document}

\maketitle

\begin{abstract}
We report on the detection of a rich water reservoir in the protostellar envelope of the Class\,0 source HH211. In striking contrast to all other molecules detected with Herschel/PACS, water emission peaks around the central source where both ortho and para forms are detected. The measured ortho-to-para ratio of just 0.65 indicates formation of water-ice at very low temperatures and a non-destructive photo-desorption process around the protostar. While part of the H$_2$O emission is likely related to collisional excitation, the centralized morphology around the protostar suggests that radiative excitation is also significant, despite the fact that radiation appears to have a very different impact on the H$_2$O molecules when compared to the terminal outflow shocks. The very low ortho-to-para ratio suggests that H$_2$O around the protostar originates from primordial envelope material that has never been thermally processed before. 
\end{abstract}

\firstsection 
\section{Introduction}

Observations of the abundance ratio of nuclear spin isomers of H$_2$O can possibly provide clues about the formation conditions and the thermal processing of molecules. For temperatures above 50\,K the H$_2$O ortho-to-para ratio (OPR) is $\sim$\,3 while for lower temperatures the ratio tends to zero. Therefore low OPR can be associated with the formation of H$_2$O at very low temperature environments. Processes leading to the desorption of water into the gas phase can alter the low-temperature OPR and modify it toward the thermal equilibrium value of 3. Non-destructive photo-desorption of H$_2$O is believed to preserve the original formation value of OPR (\cite{vanDishoeck:13a}), therefore providing clues about the thermal history of the molecule. We here report on the detection of an unusually low water - OPR in the envelope of HH\,211, indicating that the water originates from primordial, unprocessed envelope material.

\begin{figure}[!h]
\begin{center}
 \includegraphics[width=\textwidth]{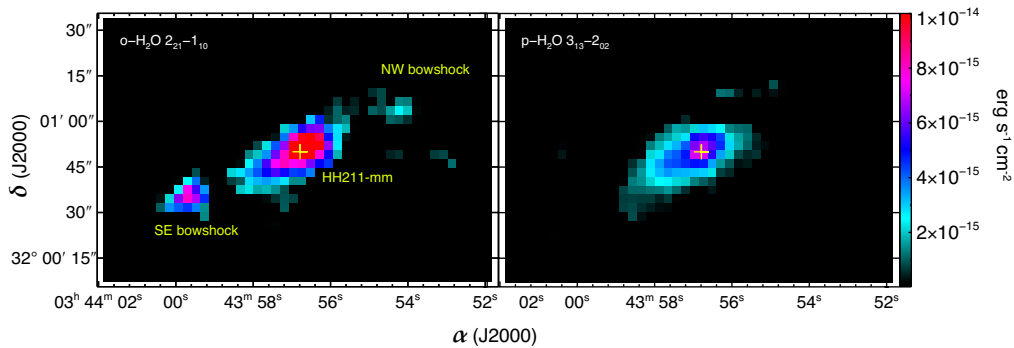}
\end{center}
 \caption{\em{Herschel/PACS maps showing examples of ortho- and para- H$_2$O transitions.}}
   \label{fig1}
\end{figure}

\section{An extremely low ortho-to-para ratio for water}
 \begin{wrapfigure}[16]{r}[0pt]{7.0cm}
\centering
\includegraphics[scale=0.32]{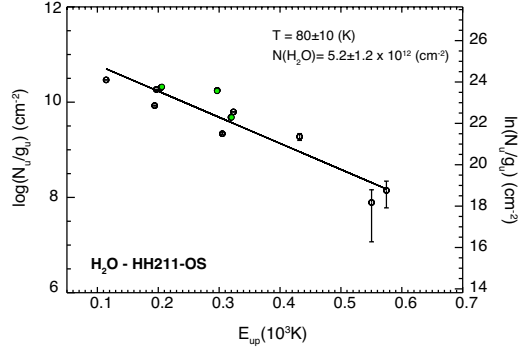}
\caption{\em{Excitation diagram for the water ortho- and para- transitions (open and filled circles, respectivelly) detected around HH\,211-mm.}}
\label{fig2}
\end{wrapfigure}
In Fig.~\ref{fig1}, we present examples of Herschel/PACS maps for the ortho- and para- forms of water (left and right panels, respectively). Water appears to be, in most cases, centralized around the protostar, as discussed extensively in \cite{Dionatos:18a}. Ortho transitions often display some weaker emission toward the SE and NW bowshocks, with the SE emission being in most cases stronger in comparison to the NW. An excitation diagram for the water emission around the protostellar source is presented in Fig.~\ref{fig2}. 
While there are just three para-H$_2$O lines detected (marked as filled circles in Fig.~\ref{fig2}), their distribution in the excitation diagram follows closely the distribution of the ortho-H$_2$O lines. Performing a separate excitation analysis, column densities are found to be $\sim$3.9 and 6.1$\times$10$^{12}$ cm$^{-2}$ for the ortho and para forms, respectively. Given the similar excitation conditions, any non-thermal excitation and opacity issues would affect the ortho and para forms in the same way, so that their column density ratio can provide a measure of their abundance ratio which we find $\sim$~0.65. The non-detection of any para-H$_2$O lines in either the SE or NW positions suggests that the OPR likely reaches the equilibrium value of 3 at the terminal shocks, as it has also been measured for H$_2$. (\cite{Dionatos:10a}).

The OPR estimated around HH211 is one of the lowest values measured in the surroundings of an embedded protostar, and is within errors comparable to the OPR$\sim$0.77 observed in the outer disk of TW Hya (\cite{Hogerheijde:11a}). It is also much lower than expected based on the temperatures of $\sim$90\,K derived from the excitation analysis. This inconsistency between the gas temperature and very low OPR values has been noticed %
%
also in the Orion PDR (\cite{Choi:14a}). Water around embedded protostars, or even pre-stellar cores, has been proposed to be the product of photodesorption from dust grains in the parental envelope (\cite{Caselli:12a}).  Given that OPR can indeed provide any information about the thermal history of the gas, then the low OPR in the surroundings of the protostar can only be interpreted through processing of primordial material that has never been thermally altered before. At the shock positions in contrast, dust temperature must have been increased multiple times through successive shocks, so that the post-shock formation and subsequent adsorption of water onto/from the dust grains reflects processing at higher temperatures.


\end{document}